\documentclass[showpacs,preprintnumbers,prb,superscriptaddress,twocolumn,reprint]{revtex4-1}
\usepackage[T1]{fontenc}
\usepackage[latin9]{inputenc}
\usepackage{textcomp}
\usepackage{amstext}
\usepackage[Symbol]{upgreek}
\usepackage{graphicx}
\usepackage{amsmath}
\usepackage{booktabs}
\usepackage{hyperref}
\usepackage{dcolumn}
\usepackage{enumerate}
\usepackage{latexsym,bm}
\makeatletter

\begin{document}
\preprint{manuscript}

\title{Sources of \emph{n}-type conductivity in GaInO$_3$}

\author{V. Wang}
\thanks{Corresponding author at: School of Science, Xi'an University of Technology, No.58, Yanxiang Road, Xi'an 710054, China, Tel./Fax: +86-29-8206-6357/6359 \\E-mail address: wangvei@icloud.com (V. Wang).}
\affiliation{Department of Applied Physics, Xi'an University of Technology, Xi'an 710054, China}  

\author{W. Xiao}
\affiliation{State Key Lab of Nonferrous Metals and Processes, General Research Institute for Nonferrous Metals, Beijing 100088, China}

\author{L.-J. Kang}
\affiliation{WPI Advanced Institute for Materials Research, Tohoku University, Sendai 980-8579, Japan}

\author{R.-J. Liu}
\affiliation{Department of Applied Physics, Xi'an University of Technology, Xi'an 710054, China}

\author{H. Mizuseki}
\affiliation{Center for Computational Science, Korea Institute of Science and Technology, Seoul 136-791,  Korea}

\author{Y. Kawazoe}
\affiliation{New Industry Creation Hatchery Center, Tohoku University, 6-6-4 Aramaki-aza-Aoba, Aoba-ku, Sendai 980-8579, Japan}
\affiliation{Institute of Thermophysics, Siberian Branch of the Russian Academy of Sciences, 1, Lavyrentyev Avenue, Novosibirsk 630090, Russia} 

\date{\today}

\begin{abstract}
Using hybrid density functional theory, we investigated formation energies and transition energies of possible donor-like defects in GaInO$_3$, with the aim of exploring the sources of the experimentally observed \emph{n}-type conductivity in this material. We predicted that O vacancies are deep donors; interstitial Ga and In are shallow donors but with rather high formation energies (>2.5 eV). Thus these intrinsic defects cannot cause high levels of \emph{n}-type conductivity. However, ubiquitous H impurities existing in samples can act as shallow donors. As for extrinsic dopants, substitutional Sn and Ge are shown to act as effective donor dopants and can give rise to highly 
\emph{n}-type conductive GaInO$_3$; while substitutional N behaviors as a compensating center. Our results provide a consistent explanation of experimental observations. 
\end{abstract}
\keywords{GaInO$_3$; hybrid density functional; \emph{n}-type conductivity; defects}
\maketitle 

\section{Introduction}
Transparent conducting oxides (TCOs) are unique materials which combine concomitant electrical conductivity and optical transparency in a single material. Thus they currently play an important role in a wide range of optoelectronic devices, such as solar cells, flat panel displays and light emitting diodes.\cite{Calnan2010,Ginley2010,Facchetti2010,Fortunato2012,Ellmer2012,Hosono2012,Martins2012,Hautier2013}  The ideal TCOs should have a carrier concentration on the order of 10$^{20}$ cm$^{-3}$ and a band-gap energy above 3.1 eV to ensure transparency to visible light. 
Recently, multicomponent oxide semiconductors have been attracting much attention as new TCOs.\cite{Hautier2010,Mizoguchi2011,Minami2013,Hautier2013}
Monoclinic GaInO$_3$ is a promising TCO due to its excellent optical transmission characteristics.\cite{Cava1994,Phillips1994,Minami1996,Minami1997} It shows a very low optical absorption coefficient on the order of a few hundreds cm$^{-1}$ which is significantly lower those of ITO, ZnO:Al and SnO$_2$:F in the visible region. It has a refractive index of around 1.65 which matches well with that of glass ($\sim$1.5). Its experimental band-gap is about 3.4 eV.
Additionally, it can be well coated on transparent substrate such as glass, fused silica, plastic, and semiconductors. Even in polycrystalline sample, the resistivity is comparable to conventional wide-band-gap transparent conductors such as indium tin oxide, while exhibiting superior light transmission. Particularly in the blue wave length region of the visible spectrum, it exhibits superior light transmission. 

The high quality \emph{n}-type GaInO$_3$ samples with conductivities of over 300 ($\Omega\cdot$cm)$^{-1}$ through doping Ge and/or Sn have been experimentally synthesized by Phillips and Minami \emph{et al.} respectively.\cite{Phillips1994,Minami1996}  They also observed that the carrier concentrations vary strongly with oxygen partial pressure \emph{p}(O$_2$) and concluded that oxygen vacancy (V$_\text{O}$) might play a key role as a native donor-like defect present in \emph{n}-type GaInO$_3$. However, it is generally accepted that V$_\text{O}$ cannot produce free electrons due to their deep donor levels even in high concentrations of V$_\text{O}$ in many TCOs, such as in ZnO,\cite{Janotti2007,Oba2008,Clark2010} SnO$_2$\cite{Janotti2011}, In$_2$O$_3$\cite{Lany2011}. In contrast, ubiquitous H impurities might be responsible for the unintentional \emph{n}-type conductivity in these materials.\cite{VandeWalle2000,Kilic2002,Hofmann2002,VandeWalle2003,Janotti2006} The structural, bonding, electronic and optical properties of GaInO$_3$ have been investigated in our previous \emph{ab-initio} studies.\cite{Wang2014} Despite extremely high \emph{n}-type conductivity in GaInO$_3$, to date, the origin mechanism of electron carriers is still unclear. 
Hence, an atomistic detailed understanding on the donor-like intrinsic and extrinsic defects possibly forming in GaInO$_3$ is necessary.

In the present work, we investigated formation energies and transition levels of intrinsic and extrinsic defects which might be responsible for the \emph{n}-type conductivity based on the hybrid density functional theory.\cite{Becke1993,Perdew1996a,Heyd2003} The recent development of hybrid density functional theory
can yield the experimental band gap values,\cite{Paier2006,Marsman2008,Park2011} and thus provides more reliable description on formation energies and transition levels of defects in semiconductors.\cite{Oba2008,Clark2010,Janotti2011,Lany2011} 
We demonstrated that (i) O vacancy and interstitial In as well as interstitial Ga are not responsible for the experimentally observed \emph{n}-type conductivity of GaInO$_3$; (ii) incorporation of H, Sn and Ge impurities act as shallow donors, which can provide a consistent explanation of experimental observations.
(iii) substitutional N on O site acts as a compensating center in \emph{n}-type GaInO$_3$.   
The remainder of this paper is organized as follows.
In Sec. II, the details of methodology and computational details are described. Sec. III presents our calculated formation energies and transition energies of various donor-like defects in GaInO$_3$. Finally, a short summary is given in Sec. IV.

\section{Methodology}
Our total energy and electronic structure calculations were carried out within a revised Heyd-Scuseria-Ernzerhof (HSE06) range-separated hybrid functional\cite{Heyd2003,Krukau2006} as implemented in VASP code.\cite{Kresse1996,Kresse1996a} In the HSE06 approach, 
the screening parameter $\mu$=0.2 {\AA}$^{-1}$ and the Hartree-Fock (HF) mixing parameter $\alpha$=28\% which means 28\% HF exchange with 72\% GGA of Perdew, Burke and Ernzerhof (PBE) \cite{Perdew1996} exchange were chosen to well reproduce the experimental band gap ($\sim$3.4 eV) of GaInO$_3$. 
The core-valence interaction was described by the frozen-core projector augmented wave (PAW) method.\cite{PAW,Kresse1999} The electronic wave functions were expanded in a plane wave basis with a cutoff of 400 eV. 
The semicore \emph{d} electrons of both Ga and In atoms were treated as core electrons. Test calculations show that the calculated formation energies of defects differ by less 0.1 eV/atom than those of the corresponding configurations in which the \emph{d} electrons were included as valence electrons. 

\begin{figure}[htbp]
\centering
\includegraphics[scale=0.25]{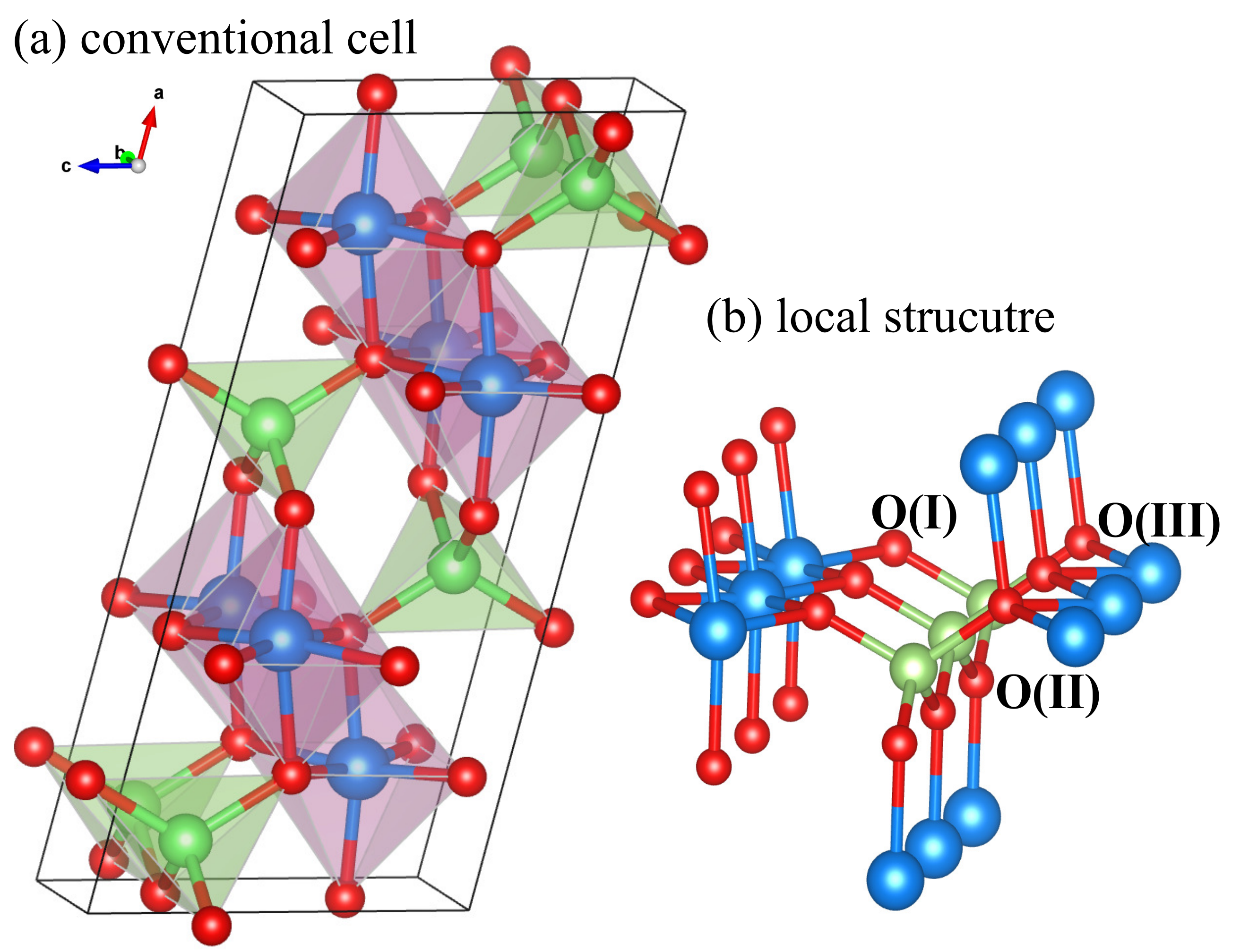}
\caption{\label{structure}(Color online) (a) Schematic polyhedral representation of GaInO$_3$ conventional cell; (b) local structure of three nonequivalent O atoms. Green, blue and red balls represent Ga, In and O atoms respectively.}
\end{figure}

As seen in Fig. \ref{structure} (a), the monoclinic GaInO$_3$ which has a c2/m space group is characterized by four lattice parameters: three vectors (\emph{a}, \emph{b} and \emph{c}) and the angle $\beta$ between \emph{a} and \emph{c} lattices.\cite{Geller1960,Cava1994,Phillips1994} Our previous studies predicted that \emph{a}, \emph{b}, \emph{c} and $\beta$ are 12.96 \AA, 3.20 {\AA}, 6.01 {\AA} and 77.89 $^\circ$ respectively, with a calculated formation energy of -8.43 eV per formula unit.\cite{Wang2014} The local structure of GaInO$_3$
is shown in Fig. \ref{structure} (b), one can find that all Ga (In) atoms site tetrahedrally (octahedrally) coordinated. There are three nonequivalent O atoms and we denote them as O(i), O(ii) and O(iii) respectively. The O(i) is threefold coordinated surrounded by two In and one Ga atoms; the O(ii) is also threefold coordinated surrounded by one In and two Ga atoms; while the O(iii) is fourfold coordinated surrounded by three In and one Ga atoms.
A more detailed discussions regarding the structural and electronic properties of GaInO$_3$ were given in our previous work.\cite{Wang2014}

The defective systems were modeled by adding (removing) an atom to (from) a 1$\times$4$\times$2 supercell consisting of 160 atoms. A 2$\times$2$\times$2 \emph{k}-point mesh within Monkhorst-Pack scheme \cite{Monkhorst1976} was applied to the Brillouin-zone integrations in total-energy calculations.
The internal coordinates in the defective supercells were relaxed to reduce the residual force on each atom to less than 0.02 eV$\cdot${\AA}$^{\text{-1}}$. All defect calculations were spin-polarized.
To investigate the source of \emph{n}-conductivity in GaInO$_3$, the intrinsic donor-like defects, including oxygen vacancy (V$_\text{O}$), interstitial Ga (Ga$_i$) and In (In$_i$) were considered in the present work.  As for extrinsic impurities, previous experimental findings have shown that substitutional Sn on In sites (Sn$_\text{In}$) and Ge on Ga sites (Ge$_\text{Ga}$) are effective \emph{n}-type dopants.\cite{Phillips1994} Additionally, the incorporation of H and N impurities were also explored since they might act as ubiquitous or purposeful impurities during the growth of GaInO$_3$. 
There are several possible interstitial sites due to the low symmetry of monoclinic structure. Here we adopted the most favorable interstitial configuration in the $\beta$-Ga$_2$O$_3$ as discussed in the previous works.\cite{Varley2010,Zacherle2013}

In the charged-defect calculations, a uniform background charge was added to keep the global charge neutrality of supercell.
The formation energy of a charged defect is defined as: \cite{Zhang1991}

\begin{equation}\label{eq1}
\begin{split}
\Delta E^f_D(\alpha,q)=E_{tot}(\alpha,q)-E_{tot}(host,0)-\sum n_{\alpha}(\mu_{\alpha}^{0}+\mu_{\alpha})  \\ 
+q(\mu_{e}+\epsilon_{v})+E_{corr}[q],
\end{split}
\end{equation}

where $E_{tot}(\alpha,q)$ and $E_{tot}(host,0)$ are the total energies of the supercells with and without defect. \emph{n}$_\alpha$ is the number of atoms of species $\alpha$ added to (\emph{n}$_\alpha$>0) or removed from (\emph{n}$_\alpha$<0) the perfect supercell to create defect. $\mu_{\alpha}^{0}$ is the atomic chemical potential of species $\alpha$ which is equal to the total energy of per atom in its most stable elemental phase, namely, $\alpha$-Ga, tetragonal-In, $\alpha$-Sn, Ge, O$_2$, H$_2$ and N$_2$.
 $\mu_{\alpha}$ is relative chemical potential referenced to the corresponding $\mu_{\alpha}^{0}$. \emph{q} is the charge state of defect and $\mu_{e}$ is electron chemical potential in reference to the host valence band maximum (VBM). Therefore, $\mu_{e}$ can vary between zero and the host band-gap \emph{E}$_g$. The final term accounts for both the alignment of the electrostatic potential between the bulk  
and defective charged supercells, as well as the finite-size effects resulting from the long-range Coulomb interaction of charged defects in a homogeneous neutralizing background, as outlined by Freysoldt \emph{et al}.\cite{Freysoldt2009}, using a calculated dielectric constant of 10.2.\cite{Wang2014}
The chemical potential $\mu_{\alpha}$ can vary from O-rich to O-poor limits depending on the growth conditions. The chemical potentials of Ga, In and O atoms are subject to their lower bounds satisfied by the constraint $\mu_{\text{Ga}}$+$\mu_{\text{In}}$+3$\mu_{\text{O}}$=$\Delta$\emph{H}$_f$(GaInO$_3$), where $\Delta$\emph{H}$_f$(GaInO$_3$) is the formation energy of GaInO$_3$. They are subject to the upper bounds $\mu_{\text{O}}$$\leq$0 (O-rich limit), $\mu_{\text{Ga}}$ $\leq$0 as well as $\mu_{\text{In}}$ $\leq$0 (O-poor limit). In addition, we examined In$_{2}$O$_3$ and Ga$_{2}$O$_3$ as limiting phases and found that they do not affect our conclusions.

For an extrinsic impurity A (A=Ga, Sn, N and N), $\mu_{\text{A}}$ is limited by the formation of its corresponding solid  (gaseous) elemental phase. Additionally, $\mu_{\text{A}}$ and $\mu_{\text{O}}$ are further limited by the formation of secondary phases A$_m$O$_n$, namely,

\begin{eqnarray}
\mu_{A}\leq0, \\ 
\mu_\text{In}+\mu_\text{Ga}+3\mu_\text{O}=\Delta H_{f}(\text{GaInO}_{3}), \\ 
m\mu_\text{A}+n\mu_\text{O}\leq\Delta H_{f}(\text{A}_{m}\text{O}_{n}).
\end{eqnarray}

We take substitutional Sn on In site as an example. To avoid the formation of secondary phase SnO$_2$, $\mu_{\text{Sn}}$+2$\mu_{\text{O}}$ $\leq$ $\Delta$\emph{H}$_f$(SnO$_2$). The O-poor limit (supposing that both In and Sn are rich) is characterized by $\mu_{\text{In}}$=0, $\mu_{\text{O}}$=$\frac{1}{3}$$\Delta$\emph{H}$_f$(GaInO$_3$), and $\mu_{\text{Sn}}$<$\Delta$\emph{H}$_f$(SnO$_2$)-$\frac{2}{3}$$\Delta$\emph{H}$_f$(GaInO$_3$) as well as $\mu_{\text{Sn}}$<0 (to avoid the segregation of $\alpha$-Sn); while the O-rich limit is characterized by $\mu_{\text{O}}$=0, $\mu_{\text{In}}$=$\Delta$\emph{H}$_f$(GaInO$_3$), and $\mu_{\text{Sn}}$<$\Delta$\emph{H}$_f$(SnO$_2$) as well as $\mu_{\text{Sn}}$<0. 
It is worth mentioning that the HSE06 calculated formation energies of these complexes depend on the HF mixing parameter $\alpha$. Our calculations show that HSE06 ($\alpha$=28\%) gives a value of -7.30 eV for $\Delta$\emph{H}$_f$(In$_2$O$_3$); while HSE06 ($\alpha$=32\%) predicts a value of -9.53 eV,\cite{Wang2013} which is quite consistent with the experimental data of -9.60 eV.\cite{Haynes2012} Thus, the available experimental formation energies of A$_m$O$_n$, together with HSE06 ($\alpha$=28\%) calculated $\mu_{\alpha}^{0}$ were adopted to determine the stability of various defects. In other words, the absolute value of the chemical potential $\mu_{\alpha}^{abs}$ is equal to the HSE06 calculated $\mu_{\alpha}$ plus the $\mu_{\alpha}^{0}$ determined from the experimental formation energies of A$_m$O$_n$.

The defect transition (ionization) energy level $\epsilon_{\alpha}$(\emph{q}/$\emph{q}^{\prime}$) is defined as the Fermi-level (\emph{E}$_\text{F}$) position for which the formation energies of these charge states are equal for the same defect,
\begin{equation}\label{eq3}
\epsilon_{\alpha}(q/q^{\prime})=[\Delta E^f_D(\alpha,q)-\Delta E^f_D(\alpha,q^{\prime})]/(q^{\prime}-q).
\end{equation}
Specifically, the defect is stable in the charge state \emph{q} when the \emph{E}$_\text{F}$ is below $\epsilon_{\alpha}(q/q^{\prime})$, while the defect is stable in the charge state q$^{\prime}$ for the \emph{E}$_\text{F}$ positions above $\epsilon_{\alpha}(q/q^{\prime})$.

\section{Results and discussion}
\begin{figure}[htbp]
\centering
\includegraphics[scale=0.38]{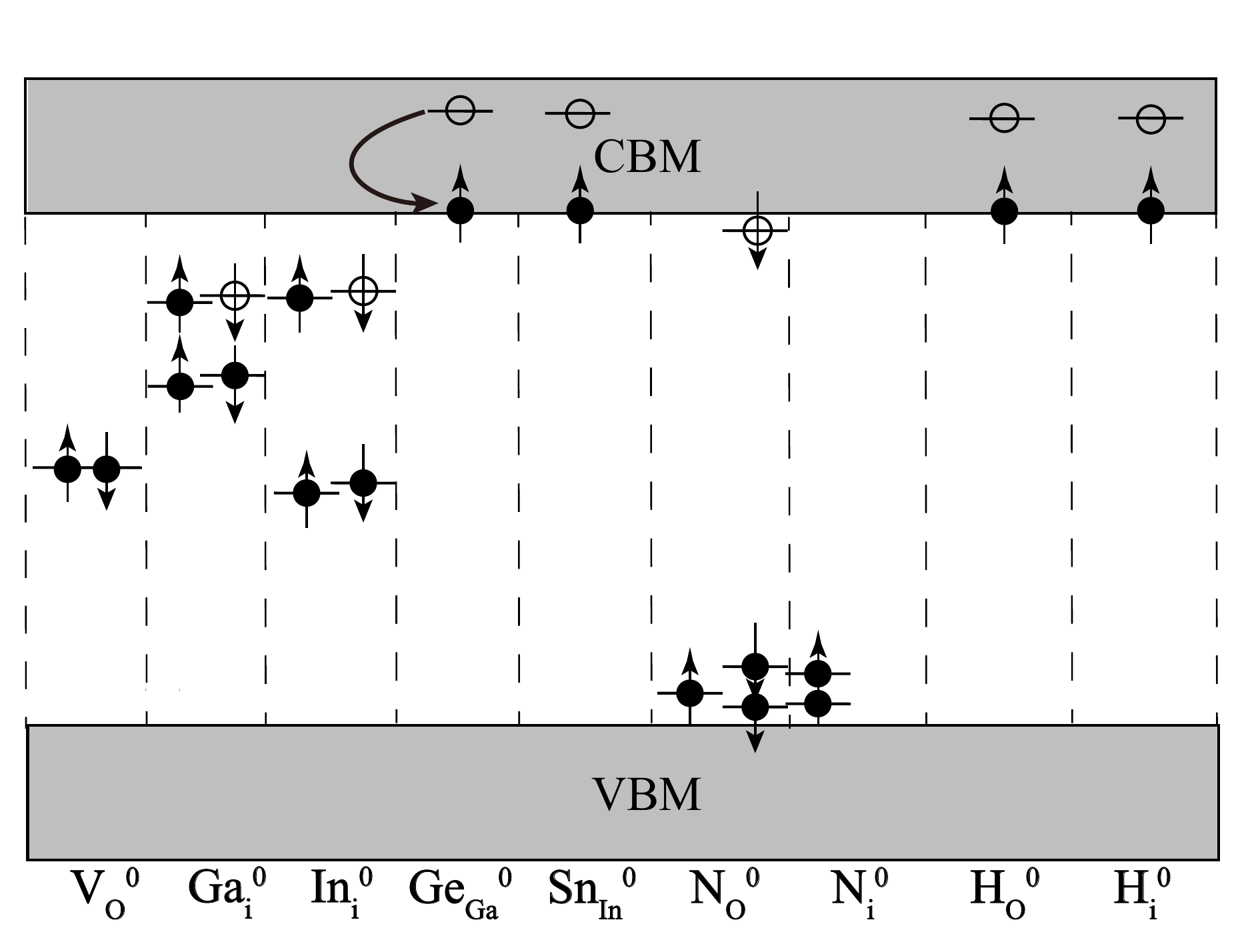}
\caption{\label{levels} Semiquantitative single particle defect levels for the neutral intrinsic and extrinsic defects in GaInO$_3$. The filled dots ($\bullet$) and open dots ($\circ$) indicate electrons and holes. The $\uparrow$ and $\downarrow$ represents spin-up and spin-down components respectively.} 
\label{fig:level}
\end{figure}

In semiconductors and insulators, the defect-levels induced by impurities or defects are either located in the band gaps, or resonant inside the continuous host bands. 
Similar to what was done in our previous studies,\cite{Wang2012,Wang2014a} a semiquantitative model which describes the single particle defect levels for all the considered neutral defects is proposed and displayed in Fig. \ref{levels}, with the aim of determining the possible charge states and sketchily catching the conductive characteristic of various defects in GaInO$_3$. One can find that V$_\text{O}$ introduces one doubly-occupied level locating around the host middle gap. Thus its possible charge states could vary from 0 to 2+, implying that V$_\text{O}$ is a donor-like defect. It is worth mentioning that the positions of defect levels would be changed over the charge state for a given defect. The local magnetic moments of V$_\text{O}^{0}$, V$_\text{O}^{1+}$ and V$_\text{O}^{2+}$ are predicted to be 0 $\mu_B$, 1 $\mu_B$ and 0 $\mu_B$ respectively, based on the filling of electrons on this defect level. This is in good agreement with the calculated findings. Considering that the resulting defect level of V$_\text{O}$ lies deep inside the band gap, V$_\text{O}$ is expected to be a deep defect and the wave functions of defect states are predicted to be localized around V$_\text{O}$ and/or its neighbors, showing an atomic-like characteristic. These speculations will be confirmed later by investigating the charge-density distribution together with transition energy levels of V$_\text{O}$.  

The neutral Ga$_i$ creates two singly-occupied levels in the spin-up component, one singly-occupied and one singly-unoccupied levels in the spin-down component. Thus, its possible charge states could range from 1- to 3+. However, it is expected that the formation of Ga$_i^{1-}$ (acting as an acceptor) is energetically unfavored as the electron affinity of Ga ion is relatively low. A similar behavior is found for In$_i$ due to the same valence electron configuration with Ga$_i$. 
From Fig. \ref{levels} we see that the neutral Ge$_\text{Ga}$, Sn$_\text{In}$, H$_\text{O}$ and H$_{i}$ introduce one singly-occupied defect level above the host conduction band minimum (CBM) independently. Since the host CBM is lower in energy than these defect levels, the electrons introduced by these impurities will drop to the CBM and occupy the perturbed conduction states. In this case, a delocalized state showing a host-band-like character is created. The system consisting of one of the above-mentioned defects has an odd number of total electrons and carries a total magnetic moments of 1 $\mu_B$.

An occupied level resonant inside the bottom of the host conduction band is the signature of a shallow donor that exhibits hydrogenic effective-mass like characteristics.\cite{Lany2005} This delocalized electron at the CBM is loosely bound to the donor whose \emph{core} is now in the charge state of 1+. We expect that Ge$_\text{Ga}^{1+}$, Sn$_\text{In}^{1+}$, H$_\text{O}^{1+}$ and H$_{i}^{1+}$ are energetically favorable when the electron chemical potential $\mu_{e}$ is below the CBM. In other words, these defect will act as donors and be stable in the 1+ charge state for all positions of the Fermi energy \emph{E}$_\text{F}$ in the band gap. The neutral N$_\text{O}$ is observed to create three singly-occupied levels above the VBM and one singly-unoccupied level just below the CBM. The neutral N$_i$ induces two singly-occupied levels in the spin-up channel above the VBM. Interestingly, both N$_\text{O}$ and N$_i$ introduce several localized states in the host forbidden bands far below the VBM (not shown in Fig. \ref{levels}). These defects states will not be further discussed as they do not contribute to the conductivity of GaInO$_3$.

\begin{figure}[htbp]
\centering
\includegraphics[scale=0.42]{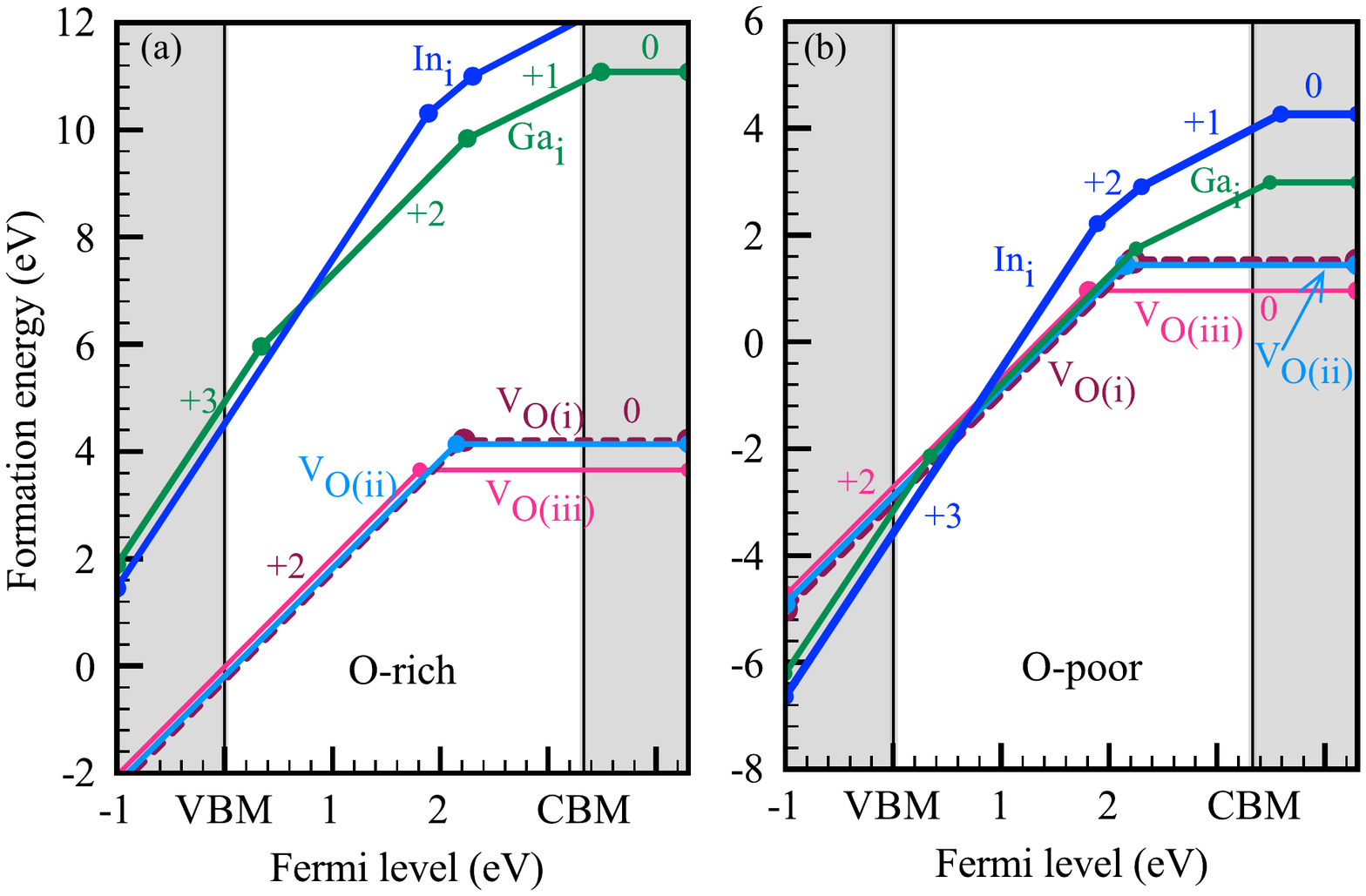}
\caption{\label{intrinsic}(Color online) Formation energies of Ga$_\text{i}$, In$_\text{i}$ and three nonequivalent O vacancies (V$_\text{O}$) in GaInO$_3$ as a function of Fermi level under (a) extreme oxygen-rich and (b) extreme oxygen-poor conditions. For the V$_\text{O}$, three nonequivalent O vacancies are labeled as V$_\text{O(i)}$, V$_\text{O(ii)}$ and V$_\text{O(iii)}$. The VBM is set to zero.}
\end{figure}

The calculated formation energies of V$_\text{O}$, Ga$_i$ and In$_i$ as a function of electron chemical potential $\mu_{e}$ are displayed in Fig. \ref{intrinsic}. For a given value of $\mu_{e}$, only the energetically stable charge state (with the lowest formation energy) of a specified defect is presented. The Fermi energies at which the slopes change correspond to the positions of thermodynamic transition levels.
One can note that the calculated transition levels $\epsilon$(2+/0) of V$_\text{O}$ are located between 1.2 and 1.6 eV below CBM depending on the site of V$_\text{O}$. 
This implies that V$_\text{O}$ acts as a deep donor.
The V$_\text{O}^{1+}$ defects are observed to be not stable for any \emph{E}$_\text{F}$ position.
The reason is attributed to the negative-U behavior which lies in the large difference in lattice relaxations between different charge states of V$_\text{O}$.
Taken as a whole, one can find that the formation energies and transition energies of oxygen vacancies on three nonequivalent O sites are lightly different due to their distinct local surroundings. The oxygen vacancy on the O(iii) site, henceforth labelled as V$_{\text{O(iii)}}$, is the most favorable configuration with a little deeper level of 1.6 eV below CBM. The behaviors of the remaining V$_{\text{O(i)}}$ and V$_{\text{O(ii)}}$ are almost indistinguishable. 
Our results on oxygen vacancies are similar to those obtained for $\beta$-Ga$_2$O$_3$.\cite{Zacherle2013}  

In contrast, both Ga$_i$ and In$_i$ act as shallow donors with $\epsilon$(1+/0) ionization energies of around 0.1 eV and 0.2 eV above CBM respectively. 
However, their formation energies are more than 2.5 eV even under \emph{n}-type conditions, in the most favorable O-poor limit. This suggests that the concentration of Ga$_i$ and In$_i$ should be negligible under equilibrium growth conditions. Based on these calculated results, we conclude that the native donor-like defects could not explain the origin of \emph{n}-type conductivity in GaInO$_3$. On the other hand, note that oxygen vacancies and cation interstitial defects are energetically stable in the 2+ and 3+ charge states respectively, with the calculated formation energies as low as -3.0 eV when the \emph{E}$_\text{F}$ is close to the VBM under O-poor growth conditions. This means that the concentrations of these native donors are high enough to certainly compensate the \emph{p}-type conductivity of GaInO$_3$ that one wants to create.
The formation of these hole-compensating defects can be suppressed by growing in the O-rich limit. More advanced experimental methods, such as nonequilibrium growth techniques may further minimize self-compensation effects in \emph{p}-type doping GaInO$_3$.

As for extrinsic impurities, previous experimental findings have shown that Sn and Ge prefer occupying the In and Ga sites respectively.\cite{Phillips1994} This is attributed to the close ionic radii of Sn (Ge) with In (Ga). The calculated shallowest transition levels are $\epsilon$(1+/0)=3.4 eV for Ge$_\text{Ga}$ and $\epsilon$(1+/0)=3.5 eV for Sn$_\text{In}$, lying just above the CBM. This means that both substitutional Sn on In sites and substitutional Ge on Ga sites are highly effective donor dopants, especially the former with a calculated formation energy of around -1.5 eV under \emph{n}-type and O-rich conditions. Our findings are in good agreement with experiments which indicate that both In and Ge can significantly enhance the \emph{n}-conductivity in GaInO$_3$ samples.\cite{Phillips1994} 

The transition energies $\epsilon$(0/1-) of substitutional N defects on three nonequivalent O sites are higher than 1.5 eV above the VBM. Clearly, N$_\text{O}$ is a deep acceptor and will not enable \emph{p}-type conductivity in GaInO$_3$. A very similar behavior has been reported for the N dopant in ZnO.\cite{Lyons2009} However, we note that N$_\text{O(i)}$ is stable in the 1- charge state with a calculated formation energy of around 0.8 eV for the \emph{E}$_\text{F}$ near the host CBM. 
It is also found that N$_\text{O}$ defects have formation energies comparable to, or even lower than those of Ge$_\text{Ga}$ and Sn$_\text{In}$ under O-rich conditions but becomes energetically less favorable under O-rich conditions. This suggests that N$_\text{O}$ can act as an electron killer and compensate the \emph{n}-type conductivity of GaInO$_3$.
This explains the experimentally observed the decrease trend on the electronic conductivity when the samples were annealed in nitrogen partial pressure.  
By comparison, N$_i$ is always energetically stable in the neutral state with a rather high formation energy of 6.5 eV, regardless of the position of the \emph{E}$_\text{F}$. This indicates that N$_i$ is electrically inactive and its concentration should be negligible under equilibrium conditions. 

\begin{figure}[htbp]
\centering
\includegraphics[scale=0.42]{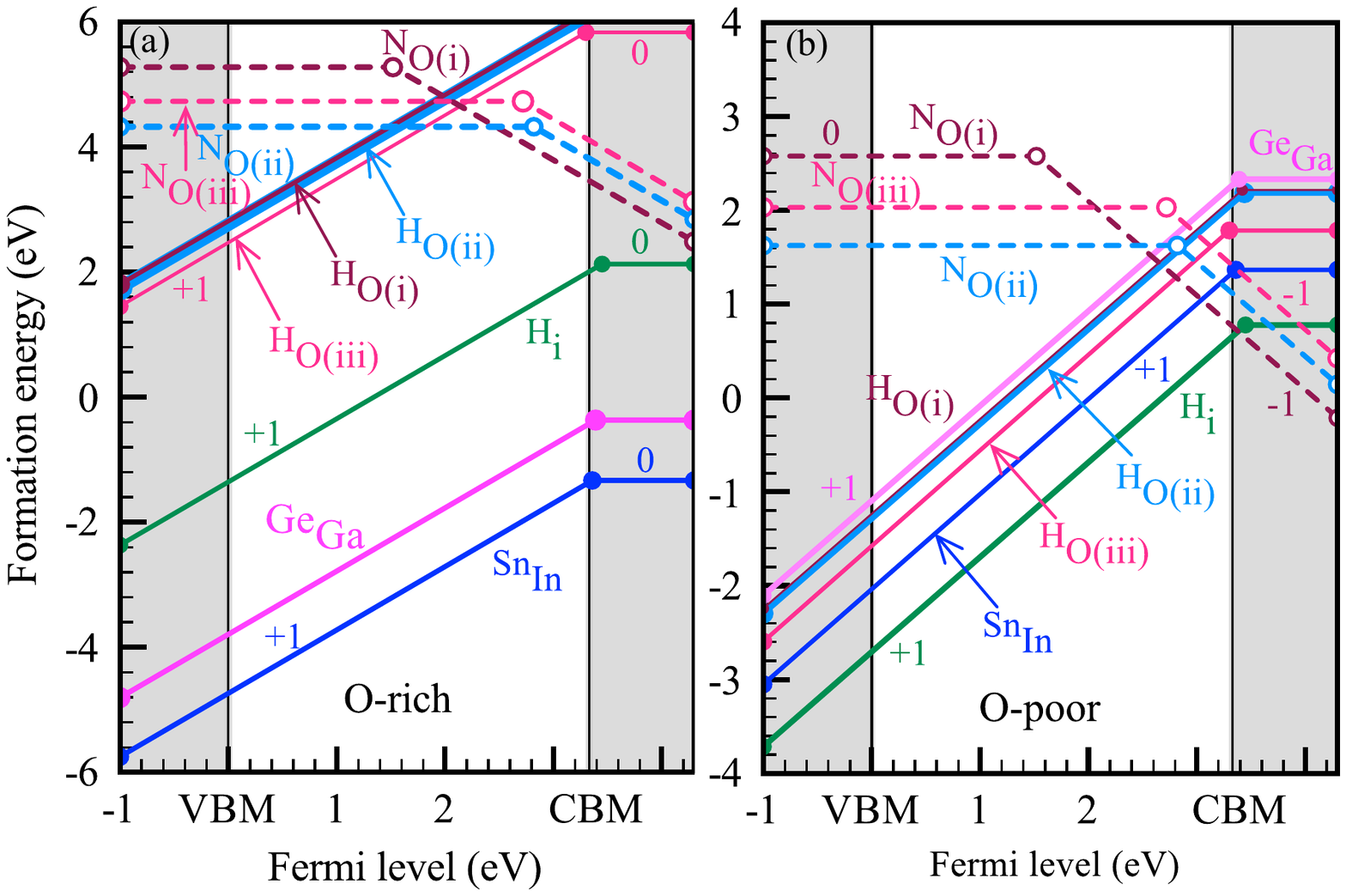}
\caption{\label{extrinsic}(Color online) Formation energies of Ge, Sn, H and N impurities in GaInO$_3$ as a function of Fermi level under (a) extreme oxygen-rich and (b) extreme oxygen-poor conditions. The VBM is set to zero.}
\end{figure}

As shown in Fig. \ref{extrinsic}, all H$_\text{O}$ defects in the different configurations act as shallow donors with the (1+/0) thermodynamic transition levels very close to the CBM. Their formation energies are lower under O-poor conditions than under O-rich ones, explaining that the concentration of H$_\text{O}$ in the samples increases with the decrease of oxygen partial pressure \emph{p}(O$_2$) during growth. \cite{Phillips1994,Minami1996} 
It is also expected that H impurities will fall into the oxygen vacancy sites, and thus H$_\text{O}$ can enhance the electrical conductivity of GaInO$_3$ under oxygen reducing (O-poor) conditions.
Besides, H$_i$ is observed to be energetically more stable than H$_\text{O}$, yielding a transition level (+1/0)=3.5 eV, just above the CBM. Hence H$_i$ also behaves exclusively as a shallow donor for any \emph{E}$_\text{F}$ value ranging from the VBM to the CBM. 
In consideration of the fact that both H$_i$ and H$_\text{O}$ serve as shallow donors, the post growth annealing in hydrogen partial pressure could help to reduce the resistivity of \emph{n}-type GaInO$_3$ samples, as was observed in experiments. Nevertheless, one can find that the 
positively charged H impurities have formation energies of less than 0.8 eV for the \emph{E}$_\text{F}$ close to VBM, implying that H impurities might act as hole compensating centers in acceptor-doped GaInO$_3$.

\begin{figure}[htbp]
\centering
\includegraphics[scale=0.38]{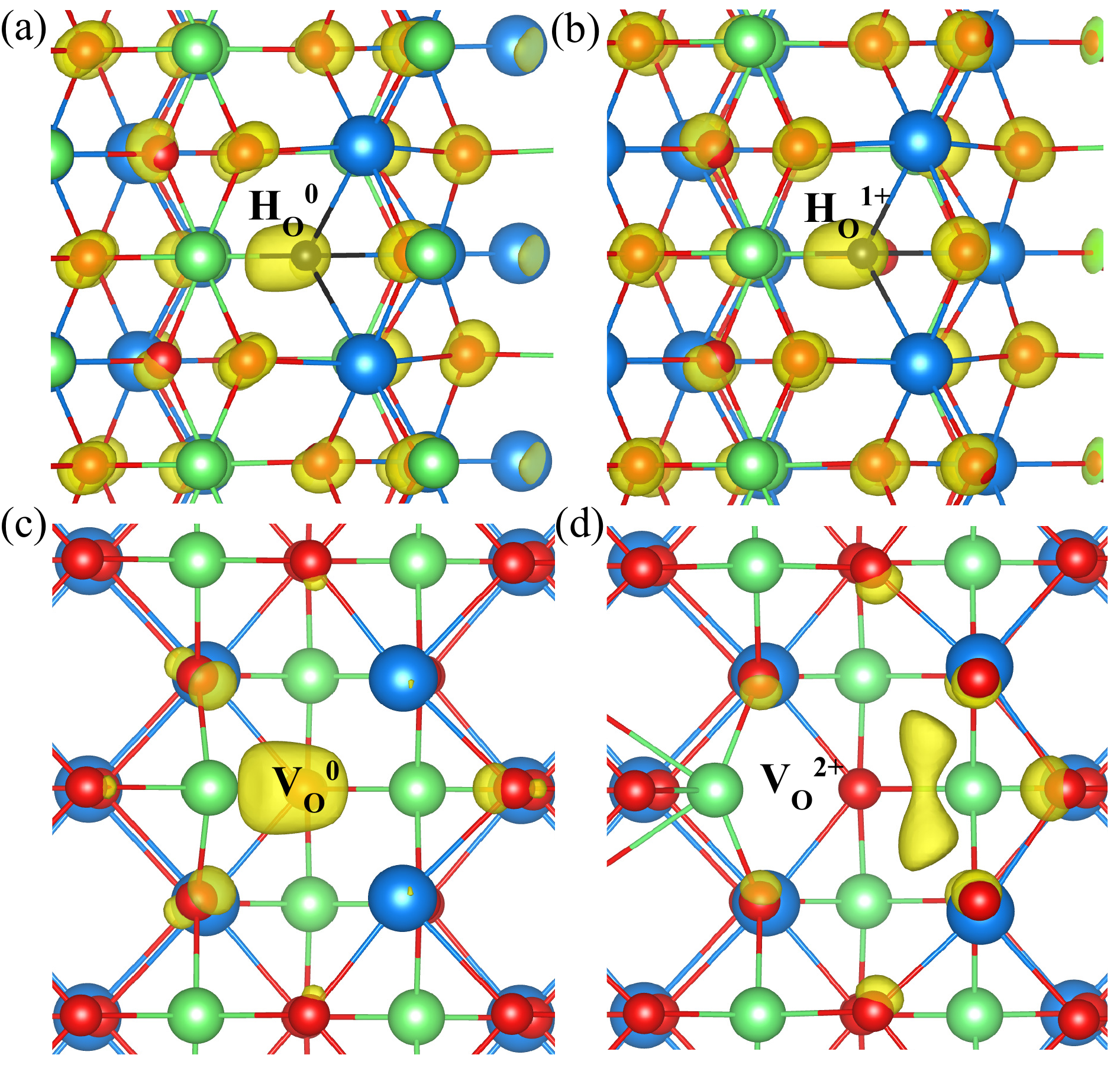}
\caption{\label{partchg}(Color online) Wavefunction squared for H$_\text{O}$ acting as a typical shallow donor in the charge states of (a) 0, (b) 1+.  Wavefunction squared for V$_\text{O}$,  an example of the typical atomic-like deep donor in the charge states of (c) 0 and (d) 2+. The charge density isosurfaces are shown at 20\% of their maximum value. Green, blue, red and black balls represent Ga, In, O and H atoms respectively.}
\end{figure}

To gain a deeper understanding of the defect states from the real space point of view, we take H$_\text{O}$ and V$_\text{O}$ as examples and plot the wavefunction squared of defect levels induced by them. As we have discussed above, H$_\text{O}$ is a typical shallow donor. From the results depicted  in Fig. \ref{partchg} we see that the wave functions of H$_\text{O}$ defect states distribute over all O atoms, showing O-2\emph{s} like and rather delocalized characteristics. The CBM of GaInO$_3$ was observed to be mainly derived from O-2\emph{s} states.\cite{Wang2014} 
This confirms that the defect level of H$_\text{O}$ is resonant 
inside the bottom of the conduction band as schematized in Fig. \ref{levels}. As expected, the wave functions of V$_\text{O}$ mainly localize at the oxygen vacancy and its seven next nearest-neighbor oxygen atoms, showing a highly-localized characteristic as V$_\text{O}$ is a deep donor. A similar behavior has been reported for V$_\text{O}$ in ZnO.\cite{Clark2010}

\begin{figure}[htbp]
\centering
\includegraphics[scale=0.42]{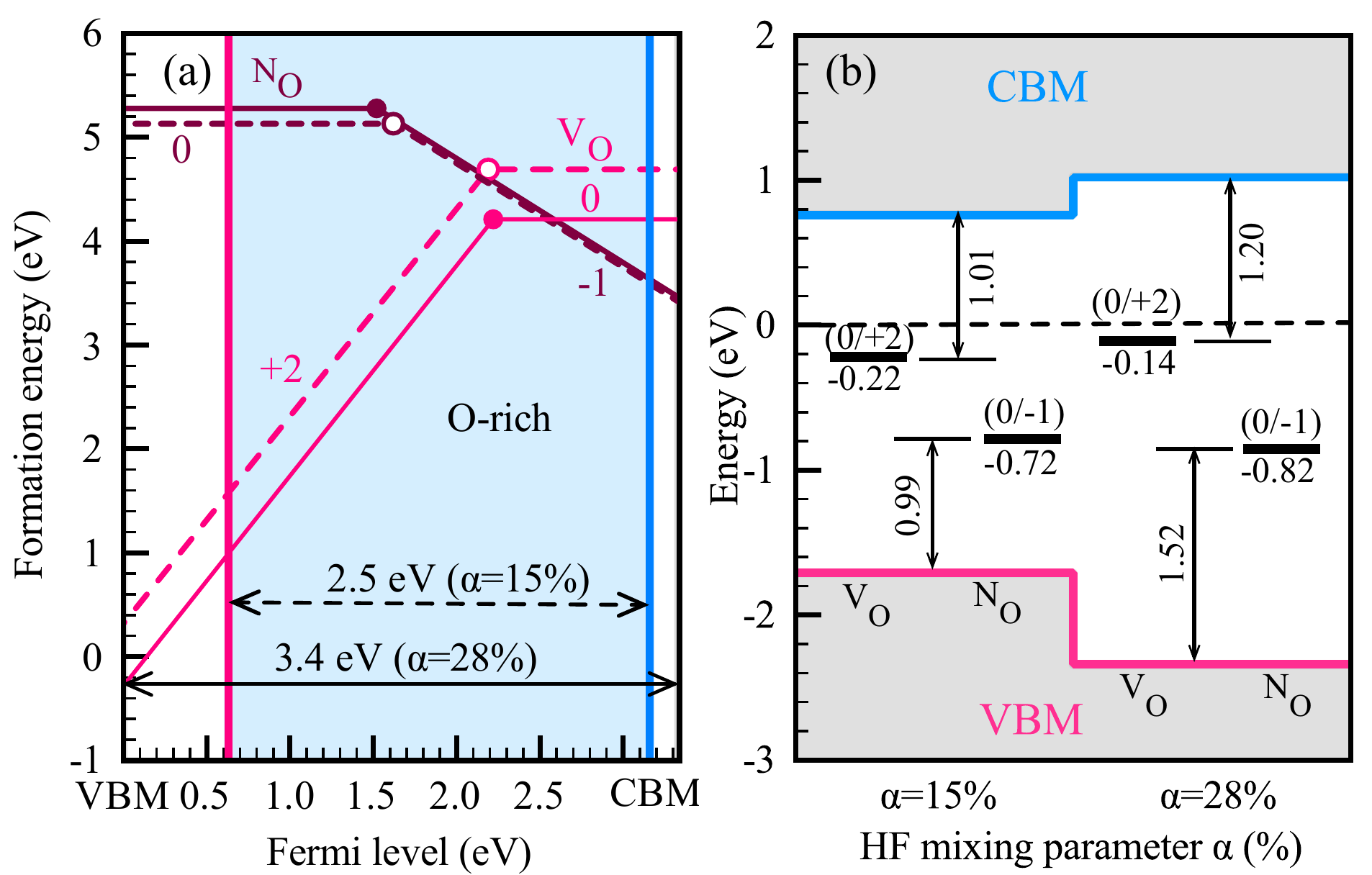}
\caption{\label{mixingparam}(Color online) (a) Formation energies of substitutional N and O vacancy as a function of Fermi level under extreme oxygen-rich condition, and (b) transition energy levels referenced to the vacuum level using HSE06 ($\alpha$=28\%) and HSE06 ($\alpha$=15\%) methods respectively. The blue region represents the HSE06 ($\alpha$=15\%) calculated band gap.}
\end{figure}

Considering that the formation energies of charged defects depend on the position of the \emph{E}$_F$ in the host band gap which is sensitive to the choice of HF mixing parameter $\alpha$, we take V$_\text{O}$ and N$_\text{O}$ as examples to investigate the roles of $\alpha$ in their stability and conductivity. From the results reported in Fig. \ref{mixingparam} (a), 
we find that the calculated formation energies of V$_\text{O}$ differ less than 0.4 eV for $\alpha$=15\% and 28\%. In contrast, the magnitude of band gap significantly decreases from 3.4 eV for $\alpha$=28\%  to 2.5 eV for $\alpha$=15\%. The underestimation of band gap using HSE06 ($\alpha$=15\%) method leads to shallower transition levels of $\epsilon$(0/1-)=0.99 eV for N$_\text{O}$ and  $\epsilon$(2+/0)=1.49 eV for V$_\text{O}$ when referred to the corresponding calculated VBM. 
To have a better understanding of the origins of these observed trends in the transition levels, we plot the transition levels on an absolute energy scale, \emph{e.g.}, referenced to the vacuum level, in Fig. \ref{mixingparam} (b). 
we note that $\alpha$ has almost negligible effects on the transition levels of V$_\text{O}$ and N$_\text{O}$ defects, with respect to the vacuum level.
Nevertheless, the host VBM moves upward, while the CBM moves downward. Consequently, the host band gap decreases along with $\alpha$. In addition, the magnitude of band offset on valence band is found to be more significantly than that on conduction band. This implies that the transition levels of acceptors should be more sensitive to the choice of $\alpha$ than those of donors in GaInO$_3$. 
In a word, the transition levels of acceptors and donors become shallow when reducing the value of $\alpha$ from 28\% to 15\%.
The rigid shifts of the host VBM and CBM are primarily responsible for the shallower transition levels which are calculated by using HSE06 ($\alpha$=15\%) approach.

\section{Summary}
In summary, we performed first-principles calculations based on hybrid density functional theory to
systematically explore the behaviors of possible donor-like defects which are ubiquitous or deliberately incorporated into GaInO$_3$ during the synthesis processes. We found that the native defects including O vacancies and interstitial Ga as well as interstitial In cannot contribute to the \emph{n}-type conductivity of GaInO$_3$ as they are either deep donors or have negligible concentrations. In contrast, our results suggest that Ge, Sn and H impurities act as shallow donors with low formation energies and they are most likely the sources of the \emph{n}-type conductivity observed in the experiments; while substitutional N acts as a compensating center in \emph{n}-type GaInO$_3$.   

\begin{acknowledgments}
We thank Profs. Wen-Tong Geng and Yu-Jun Zhao for providing valuable suggestions. Dr. Wang acknowledges the support of the Natural Science Foundation of Shaanxi Province, China (Grant No. 2013JQ1021). Prof. Kawazoe is thankful to the Russian Megagrant Project No.14.B25.31.0030 ``New energy technologies and energy carriers'' for supporting the present research.
The calculations were performed on the HITACHI SR16000 supercomputer at the Institute for Materials Research of Tohoku University, Japan.
\end{acknowledgments}

\nocite{*}
\bibliographystyle{aipnum4-1}
\end{document}